\documentclass[12pt]{article}

\newcommand{\be}{\begin{equation}}
\newcommand{\ee}{\end{equation}}
\newcommand{\bi}[1]{\vspace{-3mm} \bibitem{#1}}

\usepackage{epsfig,amsmath,amssymb,graphics,graphicx} 

\begin{document}

\begin{center}

{\bf \large Power-law Spatial Dispersion from Fractional Liouville Equation} \\

\vskip 7mm
{\bf \large Vasily E. Tarasov} \\
\vskip 3mm

{\it Skobeltsyn Institute of Nuclear Physics,\\ 
Lomonosov Moscow State University, Moscow 119991, Russia} \\
{E-mail: tarasov@theory.sinp.msu.ru} \\

\begin{abstract}
A microscopic model in the framework of fractional kinetics to
describe spatial dispersion of power-law type is suggested.
The Liouville equation with the Caputo fractional derivatives 
is used to obtain the power-law dependence of 
the absolute permittivity on the wave vector. 
The fractional differential equations for electrostatic potential 
in the media with power-law spatial dispersion are derived. 
The particular solutions of these equations for the electric potential 
of point charge in this media are considered.
\end{abstract}

\end{center}



\section{Introduction}

In the macroscopic description the spatial dispersion is represented 
by non-local connection between 
the electric displacement field ${\bf D}$ and the electric field ${\bf E}$. 
The non-locality is caused by the fact that
the field ${\bf D}$ at the point ${\bf r}$ in the medium depends 
on the values of the electric fields ${\bf E}$ not only 
in a selected point ${\bf r}$, 
but also in its neighborhood points ${\bf r}^{\prime}$.
Spatial dispersion can be described as a dependence of 
the absolute permittivity tensor of 
the medium on the wave vector \cite{SR,ABR,AR-1}. 
The electric field in the media with spatial dispersion of
the power-law type is described in the recent paper \cite{AP2013}.
The spatial dispersion is a characteristic property of the plasma-like media.
The term "plasma-like media" was introduced by Silin and Rukhadze in the book  \cite{SR}.
The plasma-like medium is characterized by the presence of free 
charge carriers, creating as they move in the medium, 
electric and magnetic fields, which significantly 
distorts the external field and the effect on the motion of 
the charges themselves \cite{SR,ABR,AR-1}.  
The plasma-like media include a wide class of object such as 
ionized gas, metals and semiconductors, 
molecular crystals and colloidal electrolytes. 
The spatial dispersion of the media leads to the set of phenomena, 
such as the rotation of the plane of polarization, anisotropy of cubic crystals 
and other \cite{AG-1,AG-2,AG-3,AR-2,KR,LL-8,Halevi,UFN-1,UFN-2,UFN-3,UFN-4,UFN-5}.

In the microscopic description of the non-local properties of the media
can be considered in the framework of models 
with long-range interactions of particles \cite{CDR,PH5,PH7}.
Equations of motion for particles with the long-range interactions in the continuous limit can give 
continuum equations with spatial derivatives of non-integer orders \cite{2006-1,2006-2,Chaos2006,LZ,CNSNS2006}.
The theory of integration and differentiation of non-integer order \cite{KST,SKM}
has a long history \cite{Ross,MKM},
and it is concerned with the names of famous mathematicians 
such as Leibniz, Liouville, Riemann, Abel, Riesz, Weyl. 
The fractional derivatives and integrals are powerful tools  
to describe complex properties of media including long-term memory, 
non-locality of power-law type, and fractality \cite{PH1,PH2,PH3,PH4,PH5,PH6,PH7,PH8,PH9,IJMPB2013}.
Using the fractional calculus we can consider different generalization
of the Liouville equation \cite{PRE2005,Chaos2006-2,MPLB2007,PH7}
that can be used in the fractional kinetics \cite{Zas,UchaikinSibatov}.

In this paper we use fractional Liouville equations to describe 
fractional kinetics for plasma-like media with
the spatial dispersion of power-law type.
The Liouville equation with the Caputo fractional derivatives 
is used to obtain the power-law dependence of 
the absolute permittivity on the wave vector. 
This allows us to have a microscopic model for 
the media with the power-law spatial dispersion, 
which are described in the recent paper \cite{AP2013}.
The appropriate fractional differential equations 
for electric potential are considered and 
particular solutions of these equations for the potential 
in the media with power-law spatial dispersion are suggested.
The difference between the point charge potential in the media with 
this type of spatial dispersion and 
the Couloumb's and Debye's potentials are described.


\section{Fractional Liouville equation}

One of the basic principles of statistical mechanics 
is the conservation of probability in the phase-space \cite{Liboff,Mart}.
The Liouville equation is an expression of the  
principle in a convenient form for the analysis.

Let us consider dynamics of system
in the phase space with dimensionless 
coordinates $({\bf x},{\bf p})=(x_1,...,x_n,p_1,...,p_n)$.
The function $\rho (t,{\bf x},{\bf p})$ 
describes probability density to find a system in 
the phase volume $d^n{\bf x} d^n{\bf p}$.
The evolution of $\rho=\rho(t,{\bf x},{\bf p})$ 
is described by the Liouville equation
\be \label{Liouv1}
\frac{\partial \rho}{\partial t} + \frac{p_i}{m} D^1_{x_i} \rho +F_i \, D^1_{p_i} \rho =0 , \ee
where $F_i=F_i({\bf x},{\bf p})$ is the force field.
Here, and later we mean the sum on the repeated index $i$ from 1 to $n$. 
Equation (\ref{Liouv1}) describes the probability conservation for
the volume element of the phase space.
If $\rho$ is the one-particle reduced distribution function, then 
the Liouville equation describes collisionless system.
Using the fractional calculus we can consider different generalization
of the Liouville equation \cite{PRE2005,Chaos2006-2,MPLB2007,PH7}
that includes derivatives of non-integer orders \cite{KST}.

We can conside a fractional generalization of the Liouville equation in the form
\be \label{Liouv2}
\frac{\partial \rho}{\partial t} + \frac{p_i}{m} \,  _0^CD^{\alpha_i}_{x_i} \rho +
F_i  \,  _0^CD^{\beta_i}_{p_i} \rho =0 ,
\ee
where we use dimensionless variables $x_i$ and $p_i$, ($i=1,...,n$). 
Here $ _0^CD^{\alpha}_{x}$ and $ _0^CD^{\beta}_{x}$ are
the Caputo fractional derivatives of order $\alpha$ and $\beta$ (see Appendix 1).

We use Caputo fractional derivatives since
a consistent formulation of fractional vector calculus,
which contains fractional differential and integral vector operations,
can be realized for Caputo differentiation 
and Riemann-Liouville integration only \cite{AP2008}.
It allows to prove the correspondent fractional generalizations of the 
Green's, Stokes' and Gauss's theorems \cite{AP2008}.
The main distinguishing feature of the Caputo fractional derivative is 
the form of the fractional generalization of the Newton-Leibniz formula
(see Lemma 2.22 \cite{KST}) in the usual form
\be \label{3-FTFC}
F(b)-F(a)  = _aI^{\alpha}_b \ _a^CD^{\alpha}_x F(x) , \quad (0<\alpha<1) .
\ee
The other feature of the Caputo fractional derivative is that, 
like the integer order derivative, 
the Caputo fractional derivative of a constant is zero. 

For simplification we consider the case $\alpha_i=\alpha$, and $\beta_i=1$
for all $i=1,...,n$.
The fractional Liouville equation is 
\be \label{Liouv3}
\frac{\partial \rho}{\partial t} + \frac{p_i}{m} \,  _0^CD^{\alpha}_{x_i} \rho +
F_i  \, D^1_{p_i} \rho =0 .
\ee
The Liouville equation with fractional derivatives with respect to coordinates
will be used to describe properties of nonlocal media.


\section{Permittivity of plasma-like nonlocal media}

In the absence of the force field ($F_i=0$), the Liouville equation (\ref{Liouv3}) gives
\be \label{Liouv4}
\frac{\partial \rho}{\partial t} + \frac{p_i}{m} \,  _0^CD^{\alpha}_{x_i} \rho =0 .
\ee
The solution of this equation is $\rho_0=\rho(t,{\bf x},{\bf p})$,
which is the distribution function unperturbed by the fields.

For a weak force field, we use 
the charge distribution function in the form 
\be
\rho=\rho_0+ \delta \rho ,
\ee
where $\rho_0$ is the stationary isotropic homogeneous 
distribution function  unperturbed by the fields, and
$\delta \rho$ is the change of $\rho_0$ by the fields. 
In the linear approximation with respect to field perturbation, we have
\be \label{Liouv5}
\frac{\partial \delta \rho}{\partial t} + 
\frac{p_i}{m} \, ( _0^CD^{\alpha}_{x_i} \delta \rho) +
F_i  \, D^1_{p_i} \rho_0 =0 .
\ee
If we consider plasma-like media, then the force ${\bf F}={\bf e}_i F_i$ is the Lorentz force
\be
{\bf F}= q {\bf E}(t,{\bf x}) + q [{\bf v},{\bf B}] ,
\ee
where $q$ is charge of particle moves with velocity ${\bf v}= {\bf p}/m$ 
in the presence of an electric field ${\bf E}={\bf e}_i E_i(t,{\bf x})$ and 
a magnetic field ${\bf B}$.
Here, and late we use the International System of Units (SI).

In an isotropic media, the distribution function depends only on the magnitude of 
the momentum, $\rho_0=\rho_0(|{\bf p}|)$. 
For such a function, the direction of the vector 
${\bf e}_i D^1_{p_i} \rho_0 $ is the same as that of ${\bf p} = m{\bf v}$, 
and its scalar product with $[{\bf v},{\bf B}]$ is equal to zero. 
Therefore, the magnetic field does not affect the distribution function
in the linear approximation. As a result, we have
\be \label{Liouv6}
\frac{\partial \delta \rho}{\partial t} + 
\frac{p_i}{m} \, (  _0^CD^{\alpha}_{x_i} \delta \rho) +
q E_i \, D^1_{p_i} \rho_0 =0 .
\ee

We assume that the perturbation 
(the function $\delta \rho$ and the field ${\bf E}$) are proportional to
\be \label{Eae}
\delta \rho, {\bf E} \ \sim \
E_{\alpha}[ i ({\bf k},{\bf x})^{\alpha} ] \cdot \exp \{- i \omega t \} ,
\ee
where $E_{\alpha}[z]$ is the Mittag-Leffler function \cite{KST}
\be
E_{\alpha} [z]:= \sum^{\infty}_{j=0} \frac{z^j}{\Gamma(\alpha \, j+1)} , 
\quad (z \in \mathbb{C}, \quad \alpha >0).
\ee
For $\alpha=1$ this function is exponent $E_{\alpha} [z]= \exp \{z\}$, and 
\[ E_{\alpha} [ i ({\bf k},{\bf x})^{\alpha} ] \,  \exp \{- i \omega t \}=
 \exp \{ i ({\bf k},{\bf x}) - i \omega t \}  . \]

We take the $x$-axis along ${\bf k}$. Then $k_x= |{\bf k}|$, 
$({\bf k},{\bf v}) =|{\bf k}| v_x$ and equation (\ref{Liouv6}) gives
\be \label{Liouv7}
i \, ( |{\bf k}|^{\alpha} v_x - \omega) \delta \rho + q ( E_i \, D^1_{p_i} \rho_0) =0 , 
\ee
where we use (see Lemma 2.23 in \cite{KST})
\be
\,  _0^CD^{\alpha}_{x} E_{\alpha}[ \lambda x^{\alpha}] = \lambda E_{\alpha}[ \lambda x^{\alpha}],
\quad (\alpha >0, \quad \lambda \in \mathbb{C}) .
\ee
As a result, we have
\be \label{delta-rho}
 \delta \rho = - \frac{ q ( E_i \, D^1_{p_i} \rho_0)}{i \, ( |{\bf k}|^{\alpha} v_x - \omega)} . 
\ee

In an unperturbed plasma-like media, the charge density is equal zero, 
since the media is isotropic. 
The charge density perturbed by the field are 
\be \label{rho-ch}
\rho_{charge} = q \int \delta \rho \, d^3 {\bf p} =
i q^2  \int \frac{ ( E_i \, D^1_{p_i} \rho_0)}{ |{\bf k}|^{\alpha} v_x - \omega }  \, d^3 {\bf p} ,
\ee
where $\rho_{charge}$ is the bound charge density.
The electric polarization vector {\bf P} is defined by the relations 
\be
\operatorname{div} {\bf P} = - \rho_{charge} .
\ee
Then
\be \label{Prho}
i ({\bf k},{\bf P})= - \rho_{charge} .
\ee
The polarization ${\bf P}$ defines the electric displacement field ${\bf D}$ as 
${\bf D} = \varepsilon_0 {\bf E} + {\bf P}$,
where $\varepsilon_0$ is the electric permittivity.
Let the field ${\bf E}$ be parallel to ${\bf k}$. 
Then ${\bf P}$ be parallel to ${\bf k}$, and 
\be \label{bfP}
{\bf P}= \Bigl( \varepsilon_{\parallel}  (|{\bf k}|) - \varepsilon_0 \Bigr) \, {\bf E} ,
\ee
where $\varepsilon_{\parallel} (|{\bf k}|)$ is the longitudinal permittivity.

Substitution of (\ref{rho-ch}) and (\ref{bfP}) into (\ref{Prho}) gives 
\be
(\varepsilon_{\parallel} (|{\bf k}|) - \varepsilon_0) \, ({\bf k}| , {\bf E})=
- q^2 \int \frac{  E_i \, D^1_{p_i} \rho_0 }{ |{\bf k}|^{\alpha} v_x - \omega - i 0}  \, d^3 {\bf p} .
\ee
Since we take the x-axis along the vector ${\bf k}$, then ${\bf E}=(E,0,0)$,
and $({\bf k}| , {\bf E})=|{\bf k}| E_x$, $E_i \, D^1_{p_i} \rho_0 =E_x \, D^1_{p_x} \rho_0$.
We introduce the function
\be
\rho_0(p_x) = \int \rho_0 (|{\bf p}|) dp_y dp_z .
\ee
As a result, the longitudinal permittivity can be calculated by the equation
\be \label{e1}
\varepsilon_{\parallel} (|{\bf k}|) = \varepsilon_0 - \frac{q^2}{ |{\bf k}| }
\int \frac{ D^1_{p_x} \rho_0(p_x) }{ |{\bf k}|^{\alpha} p_x/m - \omega - i 0}  \, d p_x .
\ee
For isotropic homogeneous case, we can use an equilibrium distribution $\rho_0(p_x)$.

\section{Longitudinal permittivity for Maxwell's distribution}

Let us consider a plasma-like medium with the equilibrium Maxwell's distribution
\be
\rho_0 (p_x) =\frac{N_q}{\sqrt{2 \pi m k_B T}} \exp \left( - \frac{p^2_x}{2m k_B T}\right) ,
\ee
where $k_B = 1.38065 \cdot 10^{-23} m^2·kg/(s^2·K)$ is the Boltzmann constant. Then
\be
D^1_{p_x} \rho_0(p_x) = 
-\frac{2 p_x N_q}{\sqrt{\pi} \, (2 m k_B T)^{3/2}} \exp \left( - \frac{p^2_x}{2m k_B T}\right) .
\ee
Here $N_q$ is the particles number density.

We define the variables
\be \label{zx}
z= \frac{p_x}{\sqrt{2m k_B T}} , \quad 
x= \sqrt{\frac{m}{2 k_B T} } \cdot \frac{\omega}{|{\bf k}|^{\alpha}} .
\ee

Equation (\ref{e1}) can be rewritten in the form
\be
\varepsilon_{\parallel} (|{\bf k}|) =\varepsilon_0 + \frac{q^2 N_q}{ |{\bf k}|^{1+\alpha} }
\frac{2 m}{\sqrt{\pi} (2 m k_B T)^{3/2}}
\int^{+\infty}_{-\infty} \frac{p_x}{ p_x - m \omega / |{\bf k}|^{\alpha} - i 0}  \, 
\exp \left( - \frac{p^2_x}{2m k_B T}\right) \, d p_x .
\ee
Using (\ref{zx}), we have
\be \label{eq-int}
\varepsilon_{\parallel} (|{\bf k}|) = \varepsilon_0 + \frac{q^2}{ |{\bf k}|^{1+\alpha} }
\frac{1}{ \sqrt{\pi} k_B T }
\int^{+\infty}_{-\infty} \frac{z \, e^{-z^2}}{ z - x - i 0}  \, d z .
\ee
Consider the integral of equation (\ref{eq-int}).
Using the formula
\[ \int^{+\infty}_{-\infty} \frac{f(z)}{z- i0} dz =
V.P. \int^{+\infty}_{-\infty} \frac{f(z)}{z} dz + i \pi f(0) , \]
and the relations
\[ \frac{z \, e^{-z^2}}{ z - x} = e^{-z^2}+ \frac{x e^{-z^2}}{ z - x} , \quad 
\int^{+\infty}_{-\infty} e^{-z^2} \, d z = \sqrt{\pi} , \]
we obtain
\be \label{int}
\int^{+\infty}_{-\infty} \frac{z \, e^{-z^2}}{ z - x - i 0}  \, d z = \sqrt{\pi}+
V.P. \int^{+\infty}_{-\infty} \frac{x e^{-z^2}}{ z - x}  \, d z + i \pi x e^{-x^2} .
\ee

Let us obtain the limiting expressions of (\ref{int}),
and therefore (\ref{eq-int}), for large and small $x$. 

\subsection*{The case of small $x$}

For $x \ll 1$, we use the variable $y=z-x$. Then
\be 
V.P. \int^{+\infty}_{-\infty} \frac{x e^{-z^2}}{ z - x} \, d z =
V.P. \int^{+\infty}_{-\infty} \frac{x e^{-(y+x)^2}}{y} \, d y .
\ee
Using
\[ e^{-(y+x)^2} = e^{-y^2} - 2y e^{-y^2} x +(2y^2-1)  e^{-y^2} x^2+ 
\frac{1}{6}(12y - 8y^3) e^{-y^2} x^3 + ... \ , \]
we get
\[
V.P. \int^{+\infty}_{-\infty} e^{-y^2} \Bigl( \frac{x}{y}-
2x^2 - \frac{x^3}{y} + 2x^3 y + 2x^4 - (4/3) y^2 x^4 +
... \Bigr) \, d y  = \]
\be \label{Eq-app-1}
= -2 \sqrt{\pi} x^2 + \sqrt{\pi} x^4 + ... \ ,
\ee
where we take into account that the integrals of the odd terms in $y$ are zero.

Substitution of (\ref{Eq-app-1}) and (\ref{int}) into (\ref{eq-int}) gives
\be \label{Final-11}
\varepsilon_{\parallel} (|{\bf k}|) = \varepsilon_0 + \frac{q^2 N_q}{ |{\bf k}|^{1+\alpha} }
\frac{1}{k_B T } \Bigl( 1- \frac{m \omega^2}{ k_B T \, |{\bf k}|^{2\alpha} } +
\frac{m^2 \omega^4}{ 4 k^2_B T^2 \, |{\bf k}|^{4\alpha} }  + ... \Bigr) .
\ee
As a result, we have
\be \label{Final-12}
\varepsilon_{\parallel} (|{\bf k}|) = 
\varepsilon_0 + \frac{q^2 N_q}{k_B T \, |{\bf k}|^{1+\alpha} }
- \frac{q^2 N_q m \omega^2}{ k^2_B T^2 \, |{\bf k}|^{3\alpha+1} } +
\frac{q^2 N_q m^2 \omega^4}{ 4 k^3_B T^3 \, |{\bf k}|^{5\alpha+1} } + ... \ .
\ee
The imaginary part of the permittivity is relatively small (not exponentially small), 
in this case because of the smallness of the phase volume in which 
the condition $|{\bf k}|^{\alpha} p_x/m - \omega =0$ is satisfied. 

The Debye radius of screening is equal to
\be \label{DR}
r_D = \sqrt{\frac{\varepsilon_0 k_B T}{N_q q^2}} . \ee
The Langmuir frequency for charged particle is
\be \label{LF}
 \Omega_L =\sqrt{\frac{N_q q^2}{m \varepsilon_0}} . \ee
Then the variable (\ref{zx}) can be represented in the form
\[ x=\frac{1}{r_D\, \Omega_L \, \sqrt{2}} \cdot \frac{\omega}{|{\bf k}|^{\alpha}} . \]
Note that ${\bf k}$, ${\bf r}$ and $x_i$ are dimensionless variable.

sing the Debye radius (\ref{DR}) and the Langmuir frequency (\ref{LF}), 
we rewrite (\ref{Final-12}) in the form
\be \label{Final-13}
\varepsilon_{\parallel} (|{\bf k}|)  \approx 
\varepsilon_0 + \varepsilon_0 \frac{1}{r^2_D  \, |{\bf k}|^{1+\alpha} }
- \varepsilon_0 \frac{\omega^2}{ r^4_D \, \Omega^2_L  \, |{\bf k}|^{3\alpha+1} } 
+ \varepsilon_0 \frac{\omega^4}{ 4 r^6_D \, \Omega^4_L  \, |{\bf k}|^{5\alpha+1} } .
\ee
Using (\ref{Final-13}), we can derive an equation for
the scalar potentials of electric field.

\subsection*{The case of large $x$}

For $x \gg 1$, we write
\[ V.P. \int^{+\infty}_{-\infty} \frac{x e^{-z^2}}{ z - x}  \, d z =
- \int^{+\infty}_{-\infty} \frac{e^{-z^2}}{ 1- z/x }  \, d z = \]
\be 
= - \int^{+\infty}_{-\infty} e^{-z^2} \left(1+\sum^{\infty}_{s=1} 
\left(\frac{z}{x} \right)^s \right) \, d z .
\ee
Integrals of the odd terms are zero. Then 
\be \label{Eq-app-2}
V.P. \int^{+\infty}_{-\infty} \frac{x e^{-z^2}}{ z - x}  \, d z =
- \sqrt{\pi} - \frac{\sqrt{\pi}}{2 x^2}-  \frac{3\sqrt{\pi}}{4 x^4}- ...
\quad (x \gg 1) .
\ee

Substituting (\ref{Eq-app-2}) and (\ref{int}) into (\ref{eq-int}), we get
\be \label{Final-21}
\varepsilon_{\parallel} (|{\bf k}|)  =1 -\frac{q^2 N_q}{\varepsilon_0 |{\bf k}|^{1+\alpha} }
\frac{1}{k_B T }
\Bigl(  \frac{k_B T}{m \omega^2} \, |{\bf k}|^{2\alpha}  +  
\frac{3 k^2_B T^2}{m^2 \omega^4} \, |{\bf k}|^{4 \alpha}  + ... \Bigr) .
\ee
The imaginary part of $\varepsilon_{\parallel} (|{\bf k}|)$ 
is exponentially small, since in a Maxwell's distribution 
only an exponentially small part of the charged particles have 
the velocity $v_x = \omega/ |{\bf k}| >> v_T= \sqrt{k_B T/m}$,
where $v_T$ is the avarage velocity of charged particles. 

As a result, we have
\be \label{Final-22}
\varepsilon_{\parallel} (|{\bf k}|)  =  \varepsilon_0  -
\frac{q^2 N_q}{ m \omega^2} \, |{\bf k}|^{\alpha-1}  -
\frac{3 q^2 N_q k_B T}{m^2 \omega^4} \, |{\bf k}|^{3 \alpha-1}  + ... \ .
\ee

Using the Debye radius (\ref{DR}) and the Langmuir frequency (\ref{LF}), 
we rewrite (\ref{Final-22}) in the form
\be \label{Final-23}
\varepsilon_{\parallel} (|{\bf k}|) \approx  \varepsilon_0  -
\varepsilon_0 \frac{\Omega^2_L}{\omega^2} \, |{\bf k}|^{\alpha-1}  -
\varepsilon_0 \frac{3 r^2_D \Omega^4_L}{\omega^4} \, |{\bf k}|^{3 \alpha-1}  .
\ee

Using equations (\ref{Final-13}) and (\ref{Final-23}),
we can obtain the scalar potentials of electric field in power-law nonlocal media,
and then describes the difference of these potentials 
from the well-known Couloumb's and Debye's potentials.


\section{Scalar potential of electric field in nonlocal media}

In the case of a static external field sources 
can create a inhomogeneous electric field ${\bf E} (t,{\bf r}) ={\bf E} ({\bf r})$. 
The electric field in the medium to be a potential
\be \label{Pot-E}
{\bf E} ({\bf r}) = - \operatorname{grad} \Phi ({\bf r}) , 
\ee
where $\Phi ({\bf r})$ is a scalar potential of electric field.

Let us consider the 3-dimensional Fourier transform 
\be
{\bf E} ({\bf r}) = \frac{1}{(2 \pi)^3} \int_{\mathbb{R}^3}
 e^{ + i ({\bf k},{\bf r})} \, {\bf E}({\bf k}) \, d^3 {\bf k}, \quad
\Phi ({\bf r}) = \frac{1}{(2 \pi)^3} \int_{\mathbb{R}^3} 
e^{ + i ({\bf k},{\bf r})} \, \Phi_{\bf k} \, d^3 {\bf k} .
\ee
The relation (\ref{Pot-E}) gives
\be \label{E=kP}
{\bf E}({\bf k}) = - i {\bf k} \, \Phi_{\bf k} .
\ee

Substituting (\ref{E=kP}) into the Maxwell equation 
\be \label{FME1}
i ({\bf k},{\bf E}(\omega,{\bf k})) \, \varepsilon ({\bf k})= \rho (\omega,{\bf k}) ,
\ee
we obtain
\be \label{FME1b}
|{\bf k}|^2 \, \varepsilon_{\parallel} \, (|{\bf k}|) \, \Phi_{\bf k} = \rho_{\bf k} ,
\ee
where $\rho_{\bf k}=\rho (0,{\bf k})$.
Note that equation (\ref{FME1b}) does not contain 
the transverse permittivity $\varepsilon_{\perp}(|{\bf k}|)$.

When the field source in the medium is the resting point charge, 
the charge density is described by delta-distribution
\be \label{delta}
\rho({\bf r}) = Q \, \delta^{(3)} ({\bf r}) ,
\ee
where we have assumed that the charge is at the beginning of the coordinate system.
Therefore the electrostatic potential of the point charge in the isotropic medium
according to the equation (\ref{FME1b}) has the form
\be \label{46}
\Phi ({\bf r}) = \frac{Q}{(2 \pi)^3} 
\int_{\mathbb{R}^3} e^{ + i ({\bf k},{\bf r}-{\bf r}_0)} \, 
\frac{1}{|{\bf k}|^2 \, \varepsilon_{\parallel} (|{\bf k}|)} \, d^3 {\bf k} .
\ee
The electric potential (\ref{46}) created by a point charge $Q$ 
at a distance $|{\bf r}|$ from the charge. \\

\subsection{The case of the Coulomb potential}


If we consider only the first term in equation (\ref{Final-13}), then
\[ \varepsilon_{\parallel}  (|{\bf k}|) = \varepsilon_0 , \]
where $ \varepsilon_0$ is the vacuum permittivity
($\varepsilon_0 \approx 8.854 \, 10^{-12} F·m^{-1}$).
Substituting $\varepsilon_{\parallel}  (|{\bf k}|) = \varepsilon_0$ into (\ref{FME1b}), we obtain
\be \label{FDEB-C}
|{\bf k}|^{2} \, \Phi_{\bf k} = \frac{1}{\varepsilon_0} \rho_{\bf k} .
\ee
The inverse Fourier transform of (\ref{FDEB-C}) gives
\be \label{Eq-2}
\Delta \Phi ({\bf r}) = - \frac{1}{\varepsilon_0} \rho({\bf r}) ,
\ee
where $\Delta$, is the 3-dimensional Laplacian, for which we have
\be
{\cal F}[ \Delta f({\bf r})]({\bf k})= 
- |{\bf k}|^2 \, {\cal F}[ f({\bf r})]({\bf k}) .
\ee
As a result, the electrostatic potential of the point charge (\ref{delta}) is
\be \label{Pot-C}
\Phi ({\bf r}) = \frac{1}{4 \pi \varepsilon_0} \, \frac{Q}{|{\bf r}|} .
\ee
This is the Coulomb's form of the potential. \\

\subsection{The case of the first two terms in equation (\ref{Final-13}) with $\alpha=1$}


If we consider only the first two terms in equation (\ref{Final-13}) with $\alpha=1$, then
\be \label{Debye1}
\varepsilon_{\parallel} (|{\bf k}|) = \varepsilon_0 \Bigl(1+\frac{1}{r^2_D |{\bf k}|^2} \Bigr), 
\ee
Substituting (\ref{Debye1}) into (\ref{FME1b}), we obtain
\be \label{FDEB-D}
\Bigl( |{\bf k}|^2 + \frac{1}{r^2_D} \Bigr) \, \Phi_{\bf k}  
= \frac{1}{\varepsilon_0} \rho_{\bf k} .
\ee
The inverse Fourier transform of (\ref{FDEB-C}) gives
\be \label{Eq-2b}
\Delta \Phi ({\bf r}) - \frac{1}{r^2_D} \Phi ({\bf r}) = - \frac{1}{\varepsilon_0} \rho({\bf r}) .
\ee
As a result, we get the screened potential
of the point charge (\ref{delta}) in the Debye's form:
\be \label{Pot-D}
\Phi ({\bf r}) = \frac{1}{4 \pi \varepsilon_0} \frac{Q}{|{\bf r}|} \cdot 
\exp \Bigl( - \frac{|{\bf r}|}{r_D} \Bigr) , 
\ee
where $r_D$ is the Debye radius of screening.
Its is easy to see that the Debye's potential
differs from the Coulomb's potential by factor $C_D(|{\bf r}|) = \exp (-|{\bf r}|/r_D)$.
Debye's sphere is a region with Debye's radius, in which there is an influence of charges, 
and outside of which charges are screened. \\

\subsection{The case of the first two terms in equation (\ref{Final-13}) with $\alpha \ne 1$}


If we consider only the first two terms in equation (\ref{Final-13}) with $\alpha \ne 1$, then
the longitudinal permittivity  $\varepsilon_{\parallel} (|{\bf k}|)$ is
\be \label{peps-III}
\varepsilon_{\parallel} (|{\bf k}|) = \varepsilon_0 
\Bigl( 1 + \frac{1}{r^2_D \, |{\bf k}|^{\alpha+1} } \Bigr) . 
\ee
Substituting (\ref{peps-III}) into (\ref{FME1b}), we obtain
\be \label{FDEB-III}
\Bigl( |{\bf k}|^{2} + \frac{1}{r^2_D} \, |{\bf k}|^{1-\alpha} \Bigr) \, \Phi_{\bf k} 
= \frac{1}{\varepsilon_0} \rho_{\bf k} .
\ee
The inverse Fourier transform of (\ref{FDEB-III}) gives
\be \label{Eq-III}
- \Delta \Phi ({\bf r}) + \frac{1}{r^2_D} \, (-\Delta)^{(1-\alpha)/2} \Phi ({\bf r})
= \frac{1}{\varepsilon_0} \rho({\bf r}) ,
\ee
where $(-\Delta)^{\alpha/2}$ is the fractional Laplacian in the Riesz form (see Appendix 2).

Equation (\ref{Eq-III}) is solvable, and its particular solution (see Appendix 3) has the form
\be \label{Eq-III-2}
\Phi({\bf r})= \frac{1}{\varepsilon_0} \int_{\mathbb{R}^3} 
G_{1-\alpha,2} ({\bf r} - {\bf r}^{\prime}) \, 
\rho ({\bf r}^{\prime}) \, d^3 {\bf r}^{\prime} ,
\ee
where
\be \label{G-Eq-III}
G_{1-\alpha,2} ({\bf r}) =\frac{|{\bf r}|^{-1/2}}{(2 \pi)^{3/2}} 
\int^{\infty}_0 \left( r^{-2}_D \, \lambda^{1- \alpha} + \lambda^2 \right)^{-1} 
\lambda^{3/2} \, J_{1/2} (\lambda |{\bf r}|) \, d \lambda .
\ee
The electrostatic potential of the point charge (\ref{delta}) is
\be \label{Pot-Eq-III-2}
\Phi ({\bf r}) = \frac{1}{4 \pi \varepsilon_0} \frac{Q}{|{\bf r}|} \
 \cdot C_{1-\alpha,2} (|{\bf r}|) , \quad (0<\alpha<1) ,
\ee
where the function
\be
C_{1-\alpha,2} (|{\bf r}|) =  \sqrt{\frac{2|{\bf r}|}{\pi}} \, 
\int^{\infty}_0 \frac{ \lambda^{3/2} \, J_{1/2} (\lambda |{\bf r}|)}{ 
 r^{-2}_D \, \lambda^{1- \alpha} + \lambda^2 } \, d \lambda ] =
\frac{2}{\pi} \, \int^{\infty}_0 
\frac{ \lambda \, \sin (\lambda |{\bf r}|)}{  r^{-2}_D \, \lambda^{1-\alpha} + \lambda^2 } \, d \lambda ,
\ee
describes the difference between this potential and the Coulomb's potential (\ref{Pot-C}). 
The asymptotic behavior of $C_{\alpha, \beta}(|{\bf r}|)$
for $|{\bf r}| \to \infty$ and for $|{\bf r}| \to 0$ is described in \cite{AP2013} 
(see sections 3.3.2 and 3.3.3).

\subsection{The case of the first three terms in equation (\ref{Final-13}) with $\alpha \ne 1$}


If we consider the first three terms in the equation (\ref{Final-13}) 
with $\alpha \ne 1$, then
\be \label{peps-IV}
\varepsilon_{\parallel} (|{\bf k}|)  = \varepsilon_0 
\Bigl( 1 + \frac{1}{r^2_D   \, |{\bf k}|^{1+\alpha} }
- \frac{\omega^2}{ r^4_D \, \Omega^2_L  \, |{\bf k}|^{3\alpha+1} } \Bigr) .
\ee
Substitution of (\ref{peps-IV}) into (\ref{FME1b}) gives
\be \label{FDEB-IV}
\Bigl( |{\bf k}|^{2}  + \frac{1}{r^2_D} \, |{\bf k}|^{1-\alpha} 
- \frac{\omega^2}{ r^4_D \, \Omega^2_L} \, |{\bf k}|^{1-3\alpha} \Bigr) \, \Phi_{\bf k} 
= \frac{1}{\varepsilon_0} \rho_{\bf k} .
\ee
The inverse Fourier transform of (\ref{FDEB-IV}) gives
\be \label{Eq-IV}
- \Delta \Phi ({\bf r}) + \frac{1}{r^2_D} \, (-\Delta)^{(1-\alpha)/2} \Phi ({\bf r}) 
- \frac{\omega^2}{ r^4_D \, \Omega^2_L} \, (-\Delta)^{(1- 3 \alpha)/2} \Phi({\bf r}) 
= \frac{1}{\varepsilon_0} \rho({\bf r}) ,
\ee
where we use the fractional Laplacian $(-\Delta)^{\alpha/2}$ (see Appendix 2).

Equation (\ref{Eq-IV}) is solvable, and its particular solution (see Appendix 3) has the form
\be \label{phi-Eq-IV}
\Phi({\bf r})= \frac{1}{\varepsilon_0} \int_{\mathbb{R}^3} 
G_{1-\alpha,1-3\alpha,2}  ({\bf r} - {\bf r}^{\prime}) \, 
\rho ({\bf r}^{\prime}) \, d^3 {\bf r}^{\prime} ,
\ee
where
\be a_{1}= \frac{1}{r^2_D} , \quad
a_{2} = \frac{\omega^2}{ r^4_D \, \Omega^2_L} , \ee
and
\be \label{G-Eq-IV}
G_{1-\alpha,1-3\alpha,2} ({\bf r}) =\frac{|{\bf r}|^{-1/2}}{(2 \pi)^{3/2}} 
\int^{\infty}_0 
\frac{\lambda^{3/2} \, J_{1/2} (\lambda |{\bf r}|)}{
 a_{1} \lambda^{1- \alpha} - a_{2} \lambda^{1- 3 \alpha} + \lambda^2 }
 \, d \lambda .
\ee

The electrostatic potential of the point charge (\ref{delta}) is
\be \label{Pot-Eq-IV}
\Phi ({\bf r}) = \frac{1}{4 \pi \varepsilon_0} \frac{Q}{|{\bf r}|} \
C_{1-\alpha, 1-3\alpha, 2} (|{\bf r}|) ,
\ee
where $0<\alpha<1/3$, and the function
\be
C_{1-\alpha, 1-3\alpha, 2} (|{\bf r}|) =
\frac{2}{\pi} \, \int^{\infty}_0 
\frac{ \lambda \, \sin (\lambda |{\bf r}|)}{ a_{1} \lambda^{1-\alpha} -
a_{2} \lambda^{1- 3 \alpha} + \lambda^2 } \, d \lambda ,
\quad (0<\alpha<1/3)
\ee
describes the difference between this potential and the Coulomb's potential.

\subsection{The case of the first three terms in the equation (\ref{Final-23}) with $\alpha \ne 1$}


If we consider the first three terms in the equation (\ref{Final-23}) 
with $\alpha \ne 1$, then
\be \label{peps-V}
\varepsilon_{\parallel} (|{\bf k}|) = \varepsilon_0 
\Bigl( 1 -  \frac{\Omega^2_L}{\omega^2} \, |{\bf k}|^{\alpha-1}  -
\frac{3 r^2_D \Omega^4_L}{\omega^4} \, |{\bf k}|^{3 \alpha-1}  \Bigr).
\ee
Substitution of (\ref{peps-V}) into (\ref{FME1b}) gives
\be \label{FDEB-V}
\Bigl( |{\bf k}|^{2}  -  \frac{\Omega^2_L}{\omega^2} \, |{\bf k}|^{\alpha+1}  -
\frac{3 r^2_D \Omega^4_L}{\omega^4} \, |{\bf k}|^{3 \alpha+1} \Bigr) \, \Phi_{\bf k} 
= \frac{1}{\varepsilon_0} \rho_{\bf k} .
\ee
The inverse Fourier transform of (\ref{FDEB-V}) gives
\be \label{Eq-V}
- \Delta \Phi ({\bf r}) -  \frac{\Omega^2_L}{\omega^2} \, (-\Delta)^{(\alpha+1)/2} \Phi ({\bf r}) 
- \frac{3 r^2_D \Omega^4_L}{\omega^4} \, (-\Delta)^{(3 \alpha+1)/2} \Phi ({\bf r}) 
= \frac{1}{\varepsilon_0} \rho({\bf r}) ,
\ee
where $(-\Delta)^{\alpha/2}$ is the fractional Laplacian (see Appendix 2).

If we consider only the first two terms in equation (\ref{Final-23}) with $\alpha \ne 1$, then
we have the fractional differential equation
\be \label{Eq-VI}
- \Delta \Phi ({\bf r}) - \frac{\Omega^2_L}{\omega^2} \, (-\Delta)^{(\alpha+1)/2} \Phi ({\bf r}) 
= \frac{1}{\varepsilon_0} \rho({\bf r}) .
\ee
The electrostatic potential of the point charge (\ref{delta}) has form
\be \label{Pot-Eq-VI}
\Phi ({\bf r}) = \frac{1}{4 \pi \varepsilon_0} \frac{Q}{|{\bf r}|} \
 \cdot C_{\alpha+1,2} (|{\bf r}|) .
\ee
The function
\be
C_{\alpha+1,2} (|{\bf r}|) =
\frac{2}{\pi} \, \int^{\infty}_0 
\frac{ \lambda \, \sin (\lambda |{\bf r}|)}{ 
\lambda^2 - (\Omega^2_L / \omega^2) \, \lambda^{\alpha+1} } \, d \lambda 
\ee
describes the difference between this potential and the Coulomb's potential.
The asymptotic behavior of $C_{\alpha, \beta}(|{\bf r}|)$
for $|{\bf r}| \to \infty$ and for $|{\bf r}| \to 0$ is described in \cite{AP2013} 
(see sections 3.3.2 and 3.3.3).

\section{Conclusion}

The suggested fractional kinetics of plasma-like media
gives a microscopic model for the 
electrodynamics of continuous media 
with the power-law spatial dispersion of power-law type, 
that is considered in the recent paper \cite{AP2013}.
The fractional kinetics is based on a generalization of the
Liouville equations that includes the Caputo fractional derivatives \cite{KST}. 
Using the fractional Liouville equation we obtain 
the power-law dependence of the absolute permittivity on the wave vector. 
This dependence leads to fractional differential equations 
for electrostatic potential that includes Riesz fractional derivatives.
Particular solutions of these equations, which describe the electric potential 
of the point charge in the media with power-law spatial dispersion are suggested.

\newpage

\newpage
\section*{Appendix 1}

The Caputo fractional derivative $\ _a^CD^{\alpha}_{x}$
can be defined for functions belonging to the space $AC^n[a,b]$
of absolutely continuous functions \cite{KST}.
Let $\alpha>0$ and let $n$ be given by
$n=[\alpha]+1$ for $\alpha \not \in \mathbb{N}$, and
$n=\alpha$ for $\alpha \in \mathbb{N}$.
If $f(x) \in AC^n[a,b]$, then the Caputo fractional derivatives
exist almost everywhere on $[a,b]$.
If $\alpha \not \in \mathbb{N}$, then
\[ (\ _a^CD^{\alpha}_{x}f)(x) = (_aI^{n-\alpha}_{x} D^n f)(x) =
\frac{1}{\Gamma(n-\alpha)} \int^x_a dz \frac{D^n_z f(z)}{(x-z)^{\alpha-n+1}} , \]
where $n=[\alpha]+1$.
If $\alpha=n \in \mathbb{N}$, then
\[ (\ _a^CD^{\alpha}_{x}f)(x) = D^n_x f(x) . \]
It can be directly verified that the Caputo fractional differentiation of
the power functions $(x-a)^{\beta}$ yields power functions
\[ _a^CD^{\alpha}_{x} (x-a)^{\beta}=
\frac{\Gamma(\beta+1)}{\Gamma(\alpha+\beta+1)} (x-a)^{\beta-\alpha} , \]
where $\beta>-1$.
In particular, 
then the Caputo fractional derivatives of a constant $C$ are equal to zero:
\[  _a^CD^{\alpha}_{x} C=0 . \]
For $k=0,1,2,...,n-1$, we have
\[ _a^CD^{\alpha}_{x} (x-a)^k=0, . \]
The Mittag-Leffler function $E_{\alpha}[\lambda (x-a)^{\alpha}]$
is invariant \cite{KST} with respect to the Caputo derivatives $\ _a^CD^{\alpha}_{x}$, i.e.,
\[ _a^CD^{\alpha}_{x} E_{\alpha}[\lambda (x-a)^{\alpha}]=
\lambda  E_{\alpha}[\lambda (x-a)^{\alpha}] . \]
This means that the Mittag-Leffler function is analogous to the exponential 
for the Caputo fractional derivative \cite{KST}.

\section*{Appendix 2}

For $\alpha > 0$ and "sufficiently good" functions $f(x)$,
$x \in \mathbb{R}^n$, the Riesz fractional differentiation is defined \cite{SKM,KST} 
in terms of the Fourier transform ${\cal F}$ by
\be
(-\Delta)^{\alpha/2}_x f(x)= {\cal F}^{-1} \Bigl( |{\bf k}|^{\alpha} ({\cal F} f)({\bf k}) \Bigr) .
\ee
For $\alpha >0$, the Riesz fractional derivative $(-\Delta)^{\alpha/2}$ 
can be defined in the form of the hypersingular integral (Sec. 26 in \cite{SKM}) by
\[ (-\Delta)^{\alpha/2}_x f(x)=\frac{1}{d_n(m,\alpha)} \int_{\mathbb{R}^n}
\frac{1}{|z|^{\alpha+n}} (\Delta^m_z f)(z) \, dz , \]
where $m> \alpha$, and $(\Delta^m_z f)(z)$ is a finite difference of
order $m$ of a function $f(x)$ with a vector step $z \in \mathbb{R}^n$
and centered at the point $x \in \mathbb{R}^n$:
\[ (\Delta^m_z f)(z) =\sum^m_{j=0} (-1)^j \frac{m!}{j! \, (m-j)!}  \, f(x-jz) . \]
The constant $d_n(m,\alpha)$ is defined by
\[ d_n(m,\alpha)=\frac{\pi^{1+n/2} A_m(\alpha)}{2^{\alpha}
\Gamma(1+\alpha/2) \Gamma(n/2+\alpha/2) \sin (\pi \alpha/2)} ,  \]
where
\[ A_m(\alpha)=\sum^m_{j=0} (-1)^{j-1} \frac{m!}{j!(m-j)!} \, j^{\alpha} . \]
Note that the hypersingular integral $ (-\Delta)^{\alpha/2}_x f(x)$ does not
depend on the choice of $m>\alpha$.

If $f(x)$ belongs to the space of "sufficiently good" functions, then
the Fourier transform ${\cal F}$ of the Riesz fractional derivative
is given by
\[ ({\cal F} (-\Delta)^{\alpha/2} f)({\bf k}) = |{\bf k}|^{\alpha} ({\cal F}f)({\bf k}) . \]
This equation is valid for the Lizorkin space \cite{SKM}
and the space $C^{\infty}(\mathbb{R}^n)$ of infinitely differentiable
functions on $\mathbb{R}^n$ with compact support.

\section*{Appendix 3}

Let us consider the fractional partial differential equation
\be \label{FPDE-1}
\sum^m_{k} a_k (-\Delta)^{\alpha_k/2} \Phi ({\bf r}) + a_0 \Phi ({\bf r})= \frac{1}{\varepsilon_0} \rho({\bf r}),
\ee
where $\alpha_m > ... > \alpha_1>0$, and $a_k \in \mathbb{R}$ are constants.
Here $(-\Delta)^{\alpha_k/2}$ are the fractional Laplacians in the Riesz form.



We apply the Fourier method for solving fractional equation (\ref{FPDE-1}).
The Fourier transform of the fractional Laplacian $(-\Delta)^{\alpha/2}$ is defined by
\be \label{FFL}
{\cal F}[ (-\Delta)^{\alpha/2} f({\bf r})]({\bf k})= 
|{\bf k}|^{\alpha} \, {\cal F}[ f({\bf r})]({\bf k}) .
\ee
Applying the Fourier transform ${\cal F}$ to both sides of (\ref{FPDE-1}) and using (\ref{FFL}),
we have
\be
({\cal F} \Phi)({\bf k}) = \frac{1}{\varepsilon_0} \left( \sum^m_{k=1} a_k |{\bf k}|^{\alpha_k}+a_0 \right)^{-1} 
({\cal F} \rho)({\bf k}) .
\ee
We define the fractional analog of the Green function \cite{KST}:
\be \label{FGF}
G_{\alpha}({\bf r})= {\cal F}^{-1} \Bigl[ \left( \sum^m_{k=1} a_k |{\bf k}|^{\alpha_k}+
a_0 \right)^{-1} \Bigr] ({\bf r})=
\int_{\mathbb{R}^3} \left( \sum^m_{k=1} a_k |{\bf k}|^{\alpha_k}+a_0 \right)^{-1} \
e^{ + i ({\bf k},{\bf r}) } \, d^3 {\bf k} ,
\ee
where $\alpha=(\alpha_1,...,\alpha_m)$ - is the multi-index.

The following relation
\be \label{3-1}
\int_{\mathbb{R}^n} e^{  i ({\bf k},{\bf r}) } \, f(|{\bf k}|) \, d^n {\bf k}= 
\frac{(2 \pi)^{n/2}}{ |{\bf r}|^{(n-2)/2}} 
\int^{\infty}_0 f( \lambda) \, \lambda^{n/2} \, J_{n/2-1}(\lambda |{\bf r}|) \, d \lambda
\ee
holds (see Lemma 25.1 of \cite{SKM}) for any function $f$
such that the integral in the right-hand side of (\ref{3-1}) is convergent.
Here $J_{\nu}$ is the Bessel function of the first kind.
As a result, the Fourier transform of a radial function is also a radial function.

Using (\ref{3-1}), the Green function (\ref{FGF}) can be represented (see Theorem 5.22 in \cite{KST})
in the form of the one-dimensional integral involving the Bessel function of the first kind $J_{1/2}$:
\be \label{G-1}
G_{\alpha} ({\bf r}) =\frac{|{\bf r}|^{-1/2}}{(2 \pi)^{3/2}} 
\int^{\infty}_0 \left( \sum^m_{k=1} a_k \lambda^{\alpha_k}+a_0 \right)^{-1} 
\lambda^{3/2} \, J_{1/2} (\lambda |{\bf r}|) \, d \lambda ,
\ee
where we use $n=3$, and $\alpha=(\alpha_1,...,\alpha_m)$ - is the multi-index.

If $\alpha_m > 1$ and $A_m \ne 0$, $A_0 \ne 0$, then equation (\ref{FPDE-1}) is solvable \cite{KST}.
The solution of equation (\ref{FPDE-1}) can be represented in the form of the
convolution of the functions $G({\bf r})$ and $\rho({\bf r})$:
\be \label{phi-G}
\Phi({\bf r})= \frac{1}{\varepsilon_0} \int_{\mathbb{R}^3} G_{\alpha} ({\bf r} - {\bf r}^{\prime}) \, 
\rho ({\bf r}^{\prime}) \, d^3 {\bf r}^{\prime} ,
\ee
where the Green function $G_{\alpha}(z)$ is defined by (\ref{G-1}).


We can consider fractional partial differential equation (\ref{FPDE-1}) with $a_0=0$ and $a_1 \ne 0$, 
when $m \in \mathbb{N}$, $m \ge 1$.
If $\alpha_1< 3$, $\alpha_m > 1$, $m \ge 1$, $a_1 \ne 0$, $a_m \ne 0$, $\alpha_m > ... > \alpha_1>0$, 
then equation
\be \label{FPDE-3}
\sum^m_{k=1} a_k (-\Delta)^{\alpha_k/2} \Phi ({\bf r}) = \frac{1}{\varepsilon_0} \rho({\bf r})
\ee
is solvable (Theorem 5.23 in \cite{KST}), and its particular solution is given by
\be \label{phi-G3}
\Phi({\bf r})= \frac{1}{\varepsilon_0} \int_{\mathbb{R}^3} G_{\alpha} ({\bf r} - {\bf r}^{\prime}) \, 
\rho ({\bf r}^{\prime}) \, d^3 {\bf r}^{\prime} ,
\ee
where
\be \label{G-3}
G_{\alpha} ({\bf r}) =\frac{|{\bf r}|^{-1/2}}{(2 \pi)^{3/2}} 
\int^{\infty}_0 \left( \sum^m_{k=1} a_k \lambda^{\alpha_k} \right)^{-1} 
\lambda^{3/2} \, \sqrt{\frac{2}{\pi z}} \, \sin (z) \, d \lambda .
\ee


\end{document}